\documentclass[pre,aps,epsfig,twocolumn]{revtex4}

\usepackage{amsmath,amssymb}
\usepackage{epsfig}

\begin{document}

\title{Covariance for Conic and Wedge Complete Filling.}

\author{C.\ Rasc\'{o}n$^{\star}$ and A.\ O.\ Parry$^{\,\dagger}$\\
\vspace*{.75cm}}
\affiliation{$^\star$
Grupo Interdisciplinar de Sistemas Complejos (GISC)\\
Dpto de Matem\'{a}ticas, Universidad Carlos III de Madrid\\ \vspace*{.15cm}
28911 Legan\'{e}s (Madrid), Spain\\ 
$^\dagger$ Department of Mathematics, Imperial College London,\\
London SW7 2BZ, United Kingdom}


\begin{abstract}
Interfacial phenomena associated with fluid adsorption in two dimensional
systems has recently been shown to exhibit hidden symmetries, or
covariances, which precisely relate local adsorption properties in
different confining geometries. We show that covariance also occurs in 
three dimensional systems and is likely to be verifiable experimentally
and in Ising model simulations studies. Specifically, we study complete
wetting in wedge (W) and cone (C) geometries as bulk coexistence is approached
and show that the equilibrium mid-point heights satisfy 
$l_c (h,\alpha)=l_w(\,\frac{h}{2}\,,\alpha),$ where $h$ measures
the partial pressure and $\alpha$ is the tilt angle. This covariance
is valid for both short-ranged and long-ranged intermolecular forces
and identifies both leading and next-to-leading order critical exponents
and amplitudes in the confining geometries. Connection with capillary
condensation-like phenomena is also made.

\end{abstract}

\maketitle

The central result of density functional theory, which
underpins the modern description of inhomogeneous fluids, is that the 
external potential is a unique functional of the equilibrium 
density profile $\rho({\bf r})$. That is, $\rho({\bf r})$ is sufficient 
to completely determine both the range of the surface intermolecular forces 
and the shape of the confining walls \cite{DFT,RW}. 

In spite of this rigourous result, recent studies have highlighted 
precise connections, or covariances, between adsorption properties for 
fluids at differently shaped substrates. 
For instance, the adsorption occuring at wedges and apexes can be related 
to adsorption properties at a planar substrate in a non-trivial manner 
\cite{Cov}.
Covariance behaves like a hidden symmetry for fluid interfaces and
has a number of profound implications for the theory of wetting, including
the allowed values of critical exponents and the very structure of 
interfacial Hamiltonians \cite{NoLo}.

In this paper, we describe a new example of geometrical covariance
which relates the interfacial properties of fluids in three dimensional
wedges and cones. Unlike previous cases, the covariance we describe here
occurs off bulk coexistence, for arbitrary intermolecular forces and 
in the complete wetting regime, i.e. for zero contact angle. 
This is of particular relevance since complete wetting is readily 
accessible in the laboratory \cite{Complete}, implying that the 
existence of geometrical 
covariances for fluid interfaces could be verified experimentally for
the first time. Indeed, complete wetting in wedges has already been studied 
in the laboratory with high accuracy \cite{Mistura}. 

The main result of this paper is the covariance relation
\begin{equation}
l_c (h,\alpha)=l_w(\,\frac{h}{2}\,,\alpha),
\label{WCC}
\end{equation}
where $l_w(h,\alpha)$ and $l_c(h,\alpha)$ denote the equilibrium interfacial 
heights at the mid-point of a wedge and a cone, respectively. Here $\alpha$
represents the substrate tilt angle whilst $h$ measures the 
partial pressure difference $p_{sat}-p$. This relation is valid as 
$h\rightarrow 0$ and describes {\it all} diverging contributions to the film 
thicknesses. 
We emphasise that similar to other cases of covariance, the above relation 
exists despite significant differences in the phase transitions occuring in 
the two geometries.

To begin, we review briefly the basic theory of complete wetting \cite{Wet}. 
Consider a planar wall in contact with a bulk vapour at temperature $T$ 
and chemical potential $\mu$. We suppose that $T$ is above the wetting 
temperature $T_w$ so that as $\mu$ is increased towards saturation 
$\mu_{sat}(T)$, the thickness of the adsorbed liquid layer $l_\pi$
diverges, corresponding to zero contact angle $\theta=0$. The divergence
is characterised by the power-law $l_{\pi}(h)\sim h^{-\beta_s^{co}}$
and is accompanied by the divergence of the transverse correlation length 
$\xi_\parallel\sim h^{-\nu_\parallel^{co}}$ and interfacial roughness
$\xi_\perp\sim h^{-\nu_\perp^{co}}$.

The standard way of modelling fluctuation effects at complete wetting 
is via the effective Hamiltonian \cite{Lip}
\begin{equation}
H_{\pi}[l]=\int\!\!d{\bf x}\;\left\{\;\frac{\Sigma}{2}\;(\nabla l)^2
+\,W(l)\,\right\},
\label{Hpi}
\end{equation}
where $\Sigma$ is the surface tension of the liquid-vapour interface.
For systems with long-ranged forces, the binding potential $W(l)$
is given by
\begin{equation}
W(l)=h\,l+\frac{A}{l^p},
\end{equation}
where $A$ is a Hamaker constant and $p$ depends on the range of the
forces. For the experimentally relevant case of three dimensions,
the cases $p=2,3$ correspond to van der Waals (vdW) and retarded
vdW forces respectively. In three dimensions, fluctuation
effects are not important and a mean-field analysis suffices to 
determine the critical properties yielding $\beta_s^{co}=1/(p+1)$,
$\nu_\parallel=(p+2)/2(p+1)$ and $\xi_\perp\sim\sqrt{\ln \xi_\parallel}$. 
In two dimensions, mean-field theory breaks down for
$p>2$ and the exponents are universal $\beta_s^{co}=\nu_\perp=1/3$ and
$\nu_\parallel=2/3$ corresponding to the so-called weak fluctuation
regime \cite{Lip}.

Now consider the analogous complete wetting in a 
three dimensional wedge. The wedge is characterised by 
a tilt angle $\alpha$ so that the height of the substrate above
the horizontal (say) is $\psi(x,y)=\tan\alpha\vert x \vert$ 
(see Fig.\ 1). 
The wedge geometry enhances the adsorption so that $l_w(h,\alpha)$ 
is far greater than the thickness of the layer adsorbed at
a flat substrate $l_\pi(h)$. The divergence of $l_w$ as $h\to 0$ is
accompanied by the divergence of correlation lengths $\xi_x$, $\xi_y$
across and along the wedge, and also by the (mid-point) roughness
$\xi_\perp$. However, fluctuation effects are not particularly important 
at complete wedge wetting and all these length scales can be related
to $l_w$. In particular, $\xi_x\sim\xi_y\sim l_w$ and 
$\xi_\perp\sim\sqrt{\ln l_w}$. The dominant divergence of $l_w$
follows from simple thermodynamic considerations since the 
macroscopic meniscus must be an inscribed cylinder with radius 
$\Sigma/h$, as determined by the Laplace pressure \cite{Hauge}. 
Thus, at leading
order, $l_w\approx\Sigma(\sec\alpha-1)/h$. Similar critical properties
occur for the two dimensional wedge although for this case 
$\xi_\perp\sim\sqrt{l_w}$.

To go beyond the macroscopics and obtain a more accurate expression
for $l_w$, we resort to an effective interfacial Hamiltonian 
description \cite{Wedge}. This reveals the presence of a significant 
next-to-leading-order contribution in $l_w$ both in two and three
dimensions which itself becomes macroscopic as $h\to 0$.
For convenience, we only present details based on the
shallow wedge model,
\begin{equation}
H[l,\psi]=\int\!\!d{\bf x}\;\left\{\;\frac{\Sigma}{2}\;(\nabla l)^2
+\,W(l-\psi)\;\right\},
\label{Hw}
\end{equation}
appropriate when $\tan\alpha\approx\alpha$. Nevertheless, analysis of a 
more involved drum-head model results in identical final expressions 
which are valid for arbitrary tilt angles.
Similar to complete wetting at a flat wall, a mean-field analysis 
suffices to determine the critical properties in three dimensions
since both the wedge thickness $l_w$ and the wetting layer thickness 
$l_\pi$ are much greater than the roughness.
The Euler-Lagrange equation for the relative height 
$\eta(x)\equiv l(x)-\alpha\vert x\vert$ obtained from minimising the 
Hamiltonian is
\begin{equation}
\Sigma\,\eta''=W'(\eta),
\end{equation}
which is solved subject to the boundary conditions 
$\eta'(0^+)=-\alpha$ and $\eta'(\infty)=0$.
This equation is integrable and leads to an exact (energy-like) 
expression for the mid-point height
\begin{equation}
\Sigma\;\frac{\;\alpha^2}{2}=W(l_w)-W(l_\pi),
\label{W2}
\end{equation}
where $W(l_\pi)$ is the disjoining pressure for a wetting film.
Making use of the Gibbs-Duhem equation
$\partial W(l_\pi)/\partial h=l_\pi$ in (\ref{W2}), and
reinserting the appropriate geometrical factors for arbitrary
tilt angles, we obtain
\begin{equation}
l_w(h,\alpha)\approx\frac{\,\Sigma\;(\sec\alpha\!-\!1)}{h}+
\sec\alpha\;D_w\;l_\pi(h) +\dots,
\label{lw}
\end{equation}
where the ellipses denote {\sl vanishing} contributions in the
$h\to 0$ limit, and the critical amplitude $D_w$ takes the non-trivial value 
\begin{equation}
D_w=\frac{1}{1-\beta_s^{co}}.
\label{Dw}
\end{equation}

\begin{figure}[t]
\hspace*{-.5cm}\epsfig{file=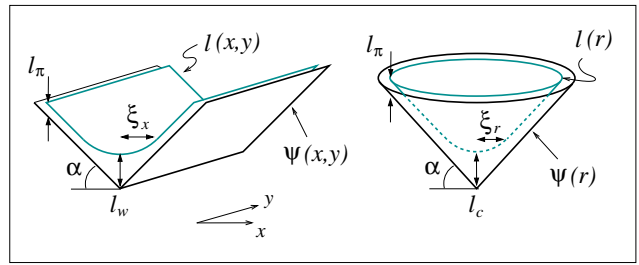,width=9.cm}
\caption{Schematic diagram of the wedge and cone geometries showing
diverging lengthscales.}
\end{figure}

Thus, the next-to-leading-order correction is directly
proportional to the {\sl planar} wetting film thickness and becomes 
macroscopic in the same limit. 
The same expansion (\ref{lw}) is also obtained in two dimensions 
from a transfer-matrix analysis and is valid in both the mean-field 
and the weak fluctuation regime.

The divergent terms in the interfacial height (\ref{lw}) contribute 
directly to the divergence of the excess adsorption in the wedge
$\Gamma$. Thus, for vdW forces, $\Gamma$ contains leading and 
next-to-leading-order contributions proportional to $h^{-2}$ and
$h^{-4/3}$ respectively. Both these contributions together with
their corresponding critical amplitudes have been measured with
remarkable accuracy in experimental studies of Ar and Kr on silicon
wedges by Mistura and co-workers \cite{Mistura}.

It is also instructive to consider the scaling properties of the
equilibrium profile. Again, for convenience, we only
consider the shallow wedge case, although the results can
be readily generalised to more acute systems.
From the Euler-Lagrange equation,
it follows that $\eta(x)=l_\pi\,\Lambda_w(x/\xi_x,x/\xi_\parallel)$
where $\xi_x=\Sigma\alpha/h$ is roughly the point of contact of
the meniscus with the wall. From this, one sees that the 
absolute interfacial height $l(x)$ divides into two regions:
a filled region for $x<\xi_x$ corresponding to the central
meniscus, and a tail for $x>\xi_x$ where the relative interfacial 
height decays towards its planar value $l_\pi$ (see Fig.\ 2).
The crossover between the regimes occurs over a distance comparable 
with the complete wetting transverse correlation length $\xi_\parallel$. 
Within the filled region, the
meniscus is, to high accuracy, parabolic with
\begin{equation}
l(x)\;\approx \;\,l_w+\,\frac{h}{2\Sigma}\;\,x^2
\label{Wm}
\end{equation}
whilst for $x>\xi_x$ the profile decays
exponentially quickly to the planar value 
\begin{equation}
l(x)\;\approx \;\,\alpha\vert x\vert \,+\; l_\pi(1+Ce^{-(x-\xi_x)/\xi_\parallel}),
\label{Wt}
\end{equation}
with $C$ a non-universal constant of order unity. These quantitative
features will be of relevance below.

We are now in position to study complete wetting in a cone geometry
and establish our main result, the covariance with wedge filling
for the mid-point heights.
The substrate height is described by the function 
$\psi(x,y)=\tan\alpha\;r$, where $r=\sqrt{x^2+y^2}$ is the radial
coordinate (see Fig.\ 1). As with the wedge, the cone strongly
enhances the adsorption compared to the planar wall and the divergence of
$l_c$ as $h\to 0$ is accompanied by the divergence of the radial
correlation length $\xi_r\sim l_c$ and mid-point roughness 
$\xi_\perp\sim\sqrt{\ln l_c}$. Thus, the transition is not fluctuation
dominated and can be understood at mean-field level.
In the cone, the macroscopic meniscus follows the surface of an
inscribed sphere of radius $2\Sigma/h$, since there are two
equal radii of curvature contributing to the Laplace pressure.
Therefore, the dominant contribution to the mid-point height is
$l_c\approx 2\Sigma(\sec\alpha-1)/h$, twice the value of the
corresponding wedge result.

A more microscopic description is afforded by the interfacial
Hamiltonian (\ref{Hw}) and its drum-head model generalisation.
Minimisation of the Hamiltonian leads to the 
Euler-Lagrange equation for the relative height 
$\eta(r)\equiv l(r)-\alpha r$,
\begin{equation}
\Sigma\,\eta''\,+\;\Sigma\;\,\frac{\eta'+\alpha}{r}\,=\; W'(\eta),
\end{equation}
which is solved subject to the same boundary conditions 
$\eta'(0)=-\alpha$ and $\eta'(\infty)=0$.
Unlike the much simpler wedge geometry, this equation is {\sl not} 
integrable and the analogue of the energy-like
equation (\ref{W2}) reads
\begin{equation}
\Sigma\;\frac{\;\alpha^2}{2}=W(l_c)-W(l_\pi)+\int_{0}^{\infty}\!\!\! dr
\;\,\frac{\eta'(\eta'\!+\!\alpha)}{r},
\label{C2}
\end{equation}
which is not an explicit expression for $l_c$.
Despite the absence of integrability, one may still
obtain the asymptotic divergence of $l_c$ analytically from
consideration of the scaling properties of the interfacial profile. 
We are not aware that analytical results for this system
have been presented before.
From the Euler-Lagrange equation, we have 
$\eta(r)=l_\pi \Lambda_c(r/\xi_r,r/\xi_\parallel)$, where
$\xi_r=2\Sigma\alpha/h$ roughly separates the meniscus region
for $r<\xi_r$ from the tail of the profile for
$r>\xi_r$. Now, within the small angle approximation, 
the meniscus is very well described by
\begin{equation}
l(r)\;\approx \;\, l_c\,+\;\frac{h}{4\Sigma}\;\,r^2,
\label{Cm}
\end{equation}
similar to the wedge result (\ref{Wm}).
The behaviour of the tails, however, is completely different for the cone.
For distances $r>\xi_r$, the profile decays {\sl algebraically}
slowly
\begin{equation}
l(r)\,\approx\;\alpha \,r \,+\;l_\pi\left(1-\frac{\xi_r}{2r}
\right)^{-\beta_s^{co}},
\label{Ct}
\end{equation}
compared with the exponential decay for the wedge (\ref{Wt}). 
It follows that there exists no rescaling of the bulk ordering field $h$
which maps the {\it full} cone profile onto the wedge profile.

Incidentally, the presence of the algebraic tails contribute
significantly to the adsorption $\Gamma$ in a cone, 
somewhat analogous to critical adsorption phenomena.
More specifically, in a cone of finite radius $R$, $\Gamma$
contains a contribution linear in $R$ arising from the tails in addition to
the usual projected area term, that scales as $R^2$.

Now focus on the determination of the mid-point conic interfacial
height $l_c$. 
The energy-like equation (\ref{C2}) for the cone is different to
that for the wedge (\ref{W2}) due to the presence of the non-integrable
final term, implying that both leading and next-to-leading order
contributions to the divergence of $l_c(h)$ which are different
to those in $l_w$, eq.\ (\ref{lw}).  The leading order divergence differs
due to a contribution $-\Sigma\alpha^2/2$ in (\ref{C2}) from the integral 
over the meniscus region $r<\xi_r$. Thus, the leading order divergence
of $l_c$ is {\it twice} that of $l_w$, consistent with the pure
macroscopic considerations above.
Similarly, the algebraic tails give rise to a contribution of order
$W(l_\pi)$ which alters the next-to-leading order singularity.
Carefully making allowance for the full scaling properties of the
profile and the specific form of the algebraic tail, it follows,
after some algebra, that $l_c$ diverges as
\begin{equation}
l_c(h,\alpha)\approx\frac{2\Sigma(\sec\alpha\!-\!1)}{h}+
\sec\alpha\;D_c\;l_\pi(h) +\dots,
\label{lc}
\end{equation}


where again the ellipses denote {\sl vanishing} contributions in the
$h\to 0$ limit. The next-to-leading-order correction term is
once more proportional to the planar wetting film thickness but
has a critical amplitude $D_c$ given by
\begin{equation}
D_c=\frac{2^{\beta_s^{co}}}{1-\beta_s^{co}},
\label{Dc}
\end{equation}
distinct from the wedge.
It is only at this point that the covariance between the
interfacial heights for the wedge and cone is manifest.
Recalling that the planar wetting film thickness grows as 
$h^{-\beta_s^{co}}$, it follows that both divergent terms
in $l_c$ mimic precisely the wedge expression (\ref{lw})
but at an effective bulk ordering field $h/2$. This leads
directly to our central result eq.\ (\ref{WCC}). 
Numerical results for the mid-point heights
obtained from minimisation of Hamiltonian (\ref{Hw})
in the wedge and cone geometries are shown in Fig.\ 2.
It can be shown that covariance also applies to the full
probability distribution functions
\begin{equation}
P_c(\,l\,;\,h\,,\alpha)=P_w(\,l\,;\,\frac{h}{2}\,,\alpha)
\end{equation}
where, in an obvious notation, $P_c(l;h,\alpha)$ denotes the PDF for the 
mid-point interfacial height in the cone, etc. By virtue of eq.\ (\ref{Wm}) 
and (\ref{Cm}), the covariance also extends to the shape of the interfacial 
profiles in the corresponding meniscus regions.
All the above statements are valid for small values of
$h$ when the mid-point interfacial height is much larger than the planar 
film thickness. 

\begin{figure}[h]
\vspace*{.35cm}\hspace*{0.2cm}\epsfig{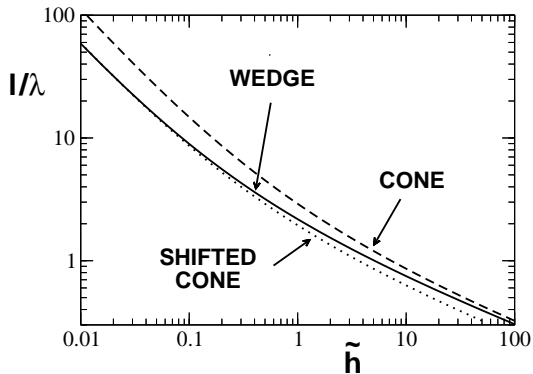}\vspace*{.15cm}\caption{Graphs of $l_w(h)$ (wedge), 
$l_c(h)$ (cone) and $l_c(2h)$
(shifted cone) obtained from numerical minimisation of the
interfacial model for vdW forces. Dimensionless units, 
where $A/\lambda^p=\Sigma\alpha^2$ and $\tilde{h}=h\lambda/\Sigma\alpha^2$
are used.}
\end{figure}

At this point, we make two remarks:

$\bullet$
Note that the nature of the next-to-leading order correction term 
in the divergence of $l_w(h)$ and $l_c(h)$ resembles Deryagin's correction 
to the Kelvin equation for capillary condensation in a slit of width $L$
\cite{Dery}. 
In the presence of thick complete wetting layers, condensation is shifted 
with respect to the bulk and occurs when the width satisfies
\begin{equation}
L\,\approx\;\frac{\Sigma}{h}\,+\;2\,D\,l_\pi(h),
\label{Ke}
\end{equation}
where Deryagin's amplitude $D$ takes the same value as (\ref{Dw}).
Only for systems with short-ranged forces, for which $\beta_s^{co}=0$,
do the right-hand sides of (\ref{lw}) and (\ref{Ke}) have a direct 
geometrical interpretation. The similarity is not coincidental
and reflects disjoining pressure contributions to the capillary 
slit, wedge and cone free energies, respectively.

$\bullet$
In addition to experimental studies, the present example of 
geometrical covariance can also be investigated in Ising model
simulations. The reason for this is that, for systems with short-ranged 
forces, the expression (\ref{lc}) for the divergence of the mid-point height
also applies to pyramidal-like geometries where the height-function is
$\psi(x,y)=\tan\alpha\,\max(|x|,|y|)$. Within the simple-cubic Ising
model, both wedge \cite{Binder} and pyramidal-like geometries
\cite{Binder2}, with $\alpha=\pi/4$,
have been used previously to study the critical filling transition
occuring at zero bulk field as $\theta\to\pi/4$. Using the same
geometries, but with stronger complete wetting surface fields (so that
$\theta=0$) and off bulk coexistence, one could readily test the 
effective Hamiltonian prediction of covariance by directly comparing 
the simulation results for the two geometries.

Finally, as stressed earlier, the covariance between the local interfacial
heights at wedge and conic filling does not contradict the central
theorem of density functional theory: the confining potential is
uniquely determined by the density profile. What covariance implies,
however, is that different confining potentials can lead to identical
local interfacial properties albeit for rescaled fields. More generally,
this implies that quite different examples of interfacial phase transitions 
share common superuniversal properties. This is perhaps more visible in the
previously discussed covariance between 2D critical wedge filling and
2D critical planar wetting. On approaching the critical filling phase
boundary, $\theta\to\alpha$, at bulk coexistence, the mid-point wedge
PDF satisfies $P_w(l;\theta,\alpha)=P_\pi(l;\theta-\alpha)$, where
$P_\pi(l;\theta)$ is the planar critical wetting PDF written as a function
of the contact angle. In these, and other examples of covariance,
the relationships between the corresponding interfacial heights in the 
different geometries only emerges after specific calculations involving
both interfacial and more microscopic models. However, the fundamental
physical origin of such relations in a more general context remains obscure.

We believe that these different covariances point towards a more
comprehensive theorem, perhaps within the framework of density functional
theory, that would enable us to further understand the relation between
geometry and fluid interfaces.

\acknowledgments{CR acknowledges support by the MCyT (Spain) 
under grant BFM2000-0004. AOP thanks the
Universidad Carlos III de Madrid for a short-stay grant.
}

\end{document}